# Symmetry guaranteed Dirac-line semimetals in two-dimensions against strong spin-orbit coupling

Deping Guo[1†], Pengjie Guo[1,2,6†], Shiliang Tan[3], Min Feng[4,5], Limin Cao[4], Zhengxin Liu[1,*], Kai Liu[1,*], Zhong-Yi Lu[1] and Wei Ji[1,*]

[1]*Beijing Key Laboratory of Optoelectronic Functional Materials & Micro-Nano Devices, Department of Physics, Renmin University of China, Beijing 100872, China*
[2]*Songshan Lake Materials Laboratory, Dongguan, Guangdong 523808, China*
[3]*Hefei National Laboratory for Physical Sciences at the Microscale, University of Science and Technology of China, Hefei, Anhui 230026, China*
[4]*School of Physics and Technology and Key Laboratory of Artificial Micro- and Nano-Structures of Ministry of Education, Wuhan University, Wuhan 430072, China*
[5]*Institute for Advanced Studies, Wuhan University, Wuhan 430072, China*
[6]*Beijing National Laboratory for Condensed Matter Physics, Institute of Physics, Chinese Academy of Sciences, Beijing 100190, China*
[†]*These authors contributed equally to this work.*
[*]*Corresponding authors. Email: liuzxphys@ruc.edu.cn (Z.L.), kliu@ruc.edu.cn (K.L.) and wji@ruc.edu.cn (W.J.);*

Several intriguing electronic phenomena and electric properties were discovered in three-dimensional Dirac nodal line semimetals (3D-DNLSM), which are, however, easy to be perturbed under strong spin-orbit coupling (SOC). While two-dimensional (2D) layers are an emerging material category with many advantages, 2D-DNLSM against SOC is yet to be uncovered. Here, we report a 2D-DNLSM in odd-atomic-layer Bi (the brick phase, another Bi allotrope), whose robustness against SOC is protected by the little co-group $C_{2v} \times Z_2^T$, the unique protecting symmetry we found in 2D. Specially, (4n+2) valence electrons fill the electronic bands in the brick phase, so that the Dirac nodal line with fourfold degeneracy locates across the Fermi level. There are almost no other low energy states close to the Fermi level; this allows to feasibly observe the neat DNLSM-induced phenomena in transport measurements without being affected by other bands. In contrast, Other VA-group elements also form the brick phases, but their DNL states are mixed with the extra states around the Fermi level. This unprecedented category of layered materials allows for exploring nearly isolated 2D-DNL states in 2D.

Topological materials, including insulators and semimetals, have attracted extensive attentions in condensed matter physics due to their rich physical properties and potential applications[1-5]. Topological insulating phases with the quantum anomalous Hall states[6] and quantum spin Hall states[7], are characterized by the fully gapped bulk and gapless edge states. The nontrivial edge states can give rise to novel quantum transport properties[4]. On the other hand, if the valence and conduction bands contact at certain symmetry-protected nodal point(s) or nodal lines in the Brillion zone, the resultant gapless phase is known as topological semimetals (TSMs) [2] and has distinctive topological properties[8-19].

The TSMs are usually categorized into Weyl, Dirac, nodal-line and other exotic semimetals[20], and are characterized by chiral anomalies[21], negative magnetoresistance[8], Fermi arcs[8-11] and drumheadlike surface states[12,22]. The TSMs exhibit tremendous novel electronic[13-15,23], optical[16-18] and magnetic behaviors[19,24]. Among these, the Dirac nodal-line semi-metals (DNLSMs) is special since it requires four-fold degeneracy in a continuous line of the BZ, which is usually fragile in the presence of SOC. Theory predicts $Cu_3PdN$[25], $Ca_3P_2$[22] and ZrSiS[26,27] as 3D DNLSMs, but they are susceptible to the perturbation of SOC, which could be approximately solved by using lighter elements, e.g. the carbon Mackay-Terrones structure[28]. Alternatively, additional symmetries can protect the nodal line of $SrIrO_3$[29] and $ReO_2$[30] even under SOC, but they are subject to compellingly experimental verification due to the mixture with certain extra states around the Fermi level. Two-dimensional van der Waals (vdW) materials, a booming category of novel materials which are easy to experimentally measure and manipulate, have many advantages in device miniaturization[31], assembly[32], interface flatness[33], and among the others. However, the corresponding symmetry operations are reduced from 3D to 2D, which further increases the difficulty of uncovering DNLSM against SOC in 2D materials.

Though 2D DNLSM materials $Cu_2Si$[34], CuSe[35] and α-Bi[36] were observed in ARPES measurement, and a category of halogen-functionalized group VA phosphorene structure was theoretically predicted[37], their nodal lines gap out under strong SOC. Here, we found a novel category of 2D intrinsic DNLSMs against SOC. Particularly, we found a novel allotrope of group VA elemental layers, i.e. the brick phase (see Figs. 1a and 1b) with the nonsymmorphic space group symmetry. A 3-atomic layer (AL) Bi thin film (see Figs. 1c and 1d) was used as a prototype to discuss the DNLSM state against SOC. Its valence electron filling guarantees the DNLSM state is near the Fermi level where other states are gapped. This exotic electronic structure hosts a ``neat'' DNLSM state, which is, to the best of our knowledge, uniquely feasible to be experimentally detected and could largely stand out the topological properties in its electrical applications. The DNLSM state also extends to 1-AL and 5-AL Bi thin films, as well as in 3-AL P, As and Sb thin films although the electronic structure of the 3-AL Bi thin film appears to be the neatest one. These 2D thin films belong to a novel category of materials that impose the nonsymmorphic symmetry group revealed in this work, which may strongly boost the research of 2D DNLSMs and their potential devices applications.

This brick phase was originally obtained in studying Bi A17 few-layers[38] which undergoes a structural transition to the brick phase if the layer thickness increases. The side view of the brick phase appears like a wall that stacked using bricks, as illustrated in Figs. 1a and 1b, and this is the reason why we call it ``brick phase''. The dashed rectangular box marks the thinnest layer of the brick phase in Fig. 1a, which is composed of three atomic-layers (ALs) of Bi atoms in an orthorhombic 2D lattice (Figs. 1c and 1d). Its lattice constants $a$ and $b$ are 4.85 and 4.53 Å, respectively, leading to Bi-Bi distances of 3.03 ($d_1$) and 3.63 Å ($d_2$) in the x-y plane. In each AL, the Bi atoms form flat zigzag chains oriented along the y axis (Fig. 1c), in different from previously found buckled chains[39-42], while the Bi atoms of the top and bottom AL show mirror symmetry with respect to the middle AL (Figs. 1a and 1d), The total energy of the 3-AL brick Bi is 91.7 meV/atom lower than the sums of a 2+1-AL (Supplementary Fig. S6). Its dynamical stability was further confirmed by our phonon dispersion spectra calculations (Supplementary Fig. S1) where the spectra exhibit no imaginary frequency, compellingly indicating the stability of the 3-AL brick phase.

The symmetry of the brick phase is described by the nonsymmorphic space group Pmma (NO. 51), whose point group G = $D_{2h} \times Z_2^T$ is generated by $C_{2x}$, $C_{2y}$, $M_z$ and $\tilde{T}$ with $\tilde{T}$=**IT**. Here, operators **I** and **T** represent the central inversion and time reversal operations, respectively, and operator {$C_{2x}$| (0, 1/2, 0)} is non-symmorphic. The spectrum degeneracy at the boundary of the Brillouin zone is protected by the irreducible projective representation (Rep) of the little co-group. For instance, the S point has a little co-group $G_0^k = D_{2h} \times Z_2^T$ whose four-dimensional projective Rep remains irreducible even if strong SOC is considered. While the combined operation $\tilde{T}$= **IT** with $\tilde{T}^2$=-1 guarantees a two-fold Kramers degeneracy in the spin degrees of freedom at any *k* point, the extra two-fold degeneracy comes from orbital degrees of freedom and is owing to the non-symmorphic nature of the symmetry operations. In order to obtain a Dirac nodal line, we need a consistent nontrivial little co-group (a subgroup of $D_{2h} \times Z_2^T$) to protect the two-fold orbital degeneracy.

We notice that the in-plane {$C_{2x}$| (0, 1/2, 0)} operation is a symmetry operation for the wave vectors along its axis, especially, the fractional translation is required to protect the orbital degeneracy. In light of this, the $C_{2v} \times Z_2^T$ group was found to be the exact little co-group symmetry in 2D satisfying the requirements of the robust four-fold degeneracy along a whole line in the BZ. This can be understood in two aspects. A), the IT symmetry with $\tilde{T}^2$=-1 protects the spin-1/2 Kramers degeneracy. The anti-commuting relationship between {$C_{2x}$| (0, 1/2, 0) and {$M_y$| (0, 1/2, 0)} is also owing to spin. B), the square of the combination of {$C_{2x}$| (0, 1/2, 0)} and IT is equal to -1, namely [{$C_{2x}$IT| (0, 1/2, 0)}]$^2$=-1 is true for all the *k* points in line ($k_x$, π, 0) of the BZ, no matter SOC is present or not. This condition cannot be satisfied in any two-dimensional matrices given that A) is unviolated. In other words, B) gives rise to an extra two-fold degeneracy owing the orbital degrees of freedom. These two factors ensure that the four-fold degeneracy in line ($k_x$, π, 0) is stable even if SOC is present (see Methods, subsection III). In addition, we also proved, using the group theory, that the dispersion relations are linear in the $k_y$, direction at the vicinity of line ($k_x$, π, 0) of the BZ (see Methods, subsection IV). In other words, our group theory analysis indicates that the

four-fold degenerated states are Dirac states and are protected by the $C_{2v} \times Z_2^T$ symmetry against SOC in the Y-S direction of the BZ.

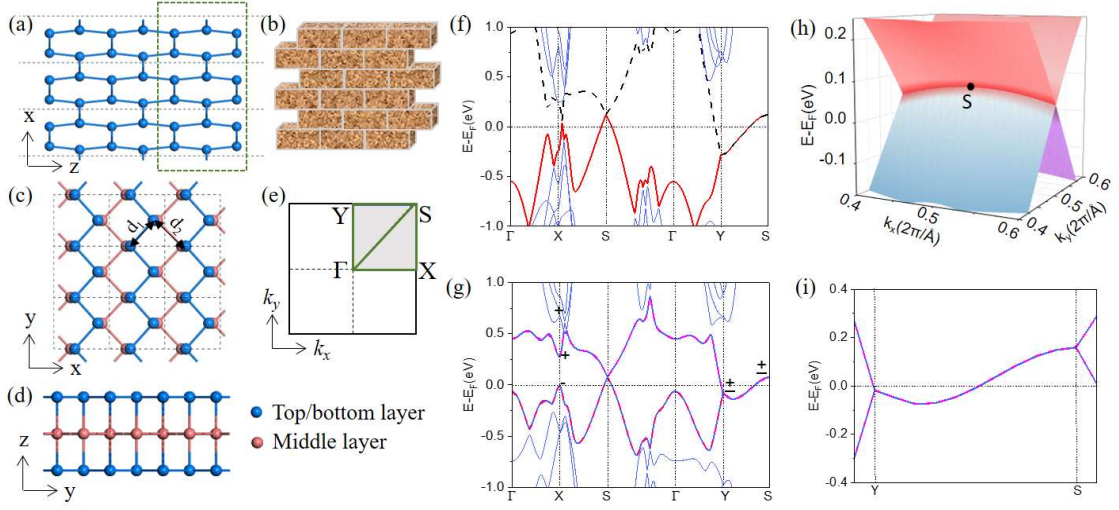

Fig. 1 (a) Side view of geometric structure of seven-layer Bi(``brick phase") and the similar brick image (b). The green dotted line in (a) is the thinnest Bi ``brick phase". (c) Top (d) side views of geometric structure of tri-layer Bi. Electronic band structures of tri-layer Bi along high-symmetry path(colored green) in Brillouin zone(e) without (f) and with (g) spin-orbit coupling. Signs ``+" and ``-" represents the symmetries of parity. (h) is the three-dimensional diagram of the linear Dirac nodal-line state of tri-layer Bi with spin-orbit coupling around S point. (i) Band structure of brick phase calculated using the HSE06 functional with SOC.

To verify our group theory analysis, we carried out density functional theory (DFT) calculations on the 3-AL Bi. Figure 1f shows the bandstructure of the 3-AL Bi without SOC along the path shown in Fig. 1e. The highest valence band (spin degenerate, red solid line) and the lowest conduction band (spin degenerate, black dashed line) contact at 0.12 eV at the S point, which forms a degenerate nodal point showing linear dispersion along the $k_y$ direction, as elucidated in Methods (subsection IV). This nodal point dispreads along the Y-S direction, forming a Dirac nodal line whose energy varying from -0.28 eV to 0.12 eV. The other state around the Fermi level is found around the time-reversal invariant point X, at which it clearly depicts a band-crossing and implies a band inversion at X.

By considering the strong SOC effect of Bi, the band structure shown in Fig. 1g explicitly indicates that the four-fold degeneracy of the nodal line in Y-S persists under SOC, which is protected by $\{C_{2x}|\ (0,\ 1/2,\ 0)\}$, $\{M_y|\ (0,\ 1/2,\ 0)\}$ and IT symmetry operations. The strong SOC effect of Bi opens a 280 meV gap at the X point, which is similar to the topological insulator gap of un-buckled bilayer bismuth at the Γ point[43,44] and leads to nontrivial edge states at the boundary of 3-AL Bi (Supplementary Fig. S2). The SOC also suppresses the dispersion of the nodal line, where the energy width reduces to 0.22 eV (from -0.14 to 0.08 eV). Both results of SOC remove the interference state at X away and substantially stand out the DNL states solely around the Fermi level, which is, to the best of our knowledge, the first report of neat DNL offering unprecedented convenience for ARPES and STM measurements.

A formula cell of the 3-AL Bi contains 30 valence electrons, which is divisible by 2 but not by 4 (the 4n+2 rule); this guarantees the DNL states are partially occupied, ensuing that the nodal line passes through the Fermi level. A 3D plot (Fig. 1h) of the DNL around point S is used to more clearly show the DNL states. The nodal line is indeed shared by the valence (blue) and conduction (red) bands. According to the group-theory analysis, the DNL states are independent of specific orbital and do not require band inversion. Orbital-resolved band structures indicate that the DNL states, developing along the Y-S direction, is linearly combined with ① and ③ or ② and ④ of the Bi $p_z$ orbitals with the negative and positive parities, forming the anti-bonding and bonding states, respectively (Supplementary Figs. S3 and S4)

The robustness of the DNL states was double checked by the HSE06 functional. The four-fold degeneracy of the DNL states maintains in the HSE06 band structure shown in Fig. 1i (with SOC) and Supplementary Fig. S5 (without SOC). It is not surprising that the quadruple degenerated DNL inevitably appears if the non-symmorphic symmetry preserves, in which the fractional-translation essentially yields a 4-dimensional irreducible projective Rep. In other words, the nonsymmorphic symmetry that we reveal guarantees the DNL, regardless the lattice constants and whether the band inversion occurs. The latter is usually required by symmorphic symmetry protected DNL semimetals [34].

Although the symmetry guarantees the neat DNL example of the Bi 3-AL, namely an electronic bandgap at the X point and the DNL solely available near the Fermi level (Fig. 2a), it is interesting to examine how in-plane strain can tune the bandgap and DNL. Figure 2b depicts the tendency of the bandwidth of the DNL ($E_1$) and the energy positions of VBM ($E_2$) and CBM ($E_3$) at the X point under uniaxial strain applied in the x direction. Energies $E_2$ and $E_3$ shift nearly linearly under external strain, slightly decreasing the size of the bandgap from the uniaxial strain of -3% to 4%. The bandwidth of the DNL, however, reduces as lattice constant $a$ shrinks. When the compression reaches -3%, a 310 meV gap is opened at the X point with VBM under the Fermi level; this indicates that the most pronounced DNL could be, most likely, detected at the -3% uniaxial strain along the x direction. (Supplementary Fig. S6)

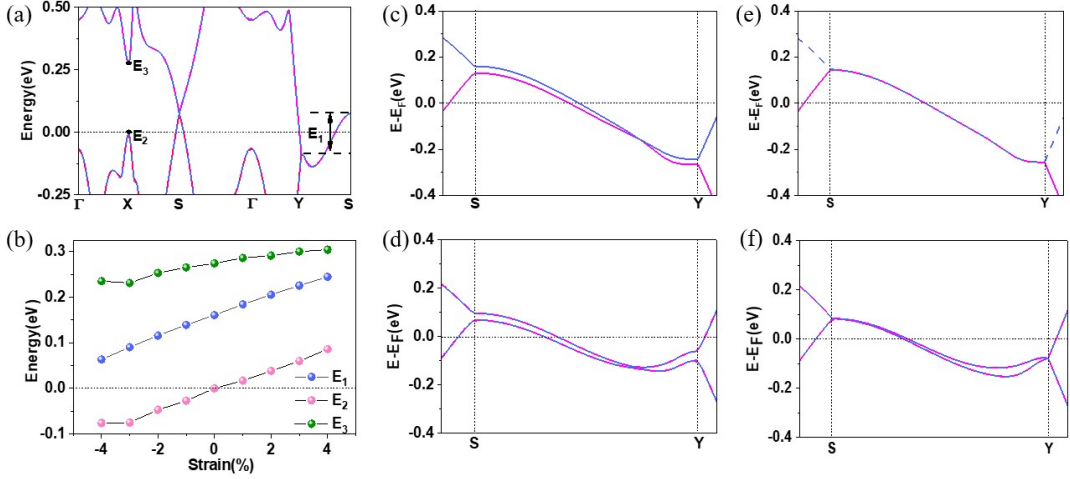

Fig. 2 (a) Definitions of width of Dirac nodal line ($E_1$), VBM ($E_2$) and CBM ($E_3$) at X point. (b) Variation of the width of Dirac nodal line, VBM and CBM at X point by applying different uniaxial strain along the x direction. Space inversion symmetry broken in Bi brick phase without (c) and with SOC (d), $C_{2x}$ symmetry with fractional translation broken in Bi brick phase without (e) and with SOC (f).

While the four-fold DNL of 3-AL brick-phase Bi is protected by the $\mathcal{C}_{2v} \times Z_2^T$ symmetry, it would be interesting to explore how the electronic bandstructure responses to breaking symmetry, e.g. degradation from DNL to Dirac point (DP) states[2]. In particular, the breaking space inversion symmetry opens a gap of the DNL regardless the inclusion of SOC (Figs. 2c and 2d). The combined operation {$C_{2x}$IT| (0, 1/2, 0)} (or, equivalently {$M_y$IT| (0, 1/2, 0)}) protects the orbital degeneracy if SOC was not considered, thus the DNL maintains even if the $C_{2x}$ symmetry with fractional translation breaks (Fig. 2e). If the SOC effect is considered, the anti-commuting relation between

{$M_y$| (0, 1/2, 0)} and {$C_{2x}$| (0, 1/2, 0)} is required to protect the orbital degeneracy (see Methods, subsection III). Breaking one of these symmetries will reduce the quadruple degeneracy into two double degeneracies, except for the high symmetry points Y and S (Fig. 2f) whose orbital degeneracy can be guaranteed by space-inversion, {My(0, 1/2, 0) and time-reversal symmetry. Therefore, the DNL, in this case, degrades into two DPs and similar DPs were also found in a 3-AL ``A7-type'' Bi structure[42].

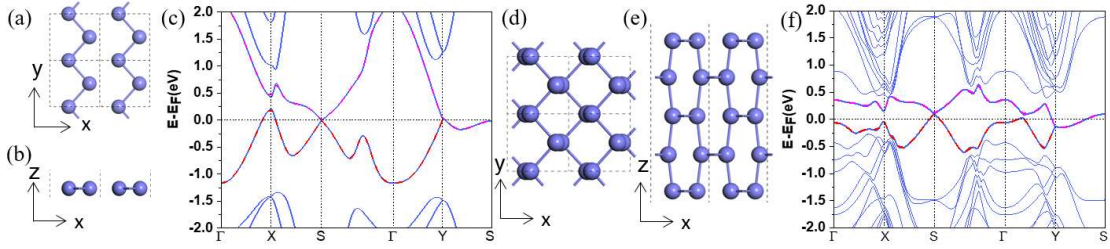

Fig. 3 (a) Top and (b) side view of geometric structure of monolayer Bi. (c) Electronic band structures of monolayer in the presence of SOC. (d) Top and (e) side view of geometric structure of five-layer Bi. (c) Electronic band structures of five-layer Bi in the presence of SOC

Although the mono-atomic layer of the brick phase Bi (Figs. 3a and 3b), most likely, needs a substrate to stabilize, we also plotted its band structure in Fig. 3c. The mono-atomic layer shares the same $C_{2v} \times Z_2^T$ symmetry with the 3-AL, which is, as expected, a DNL semimetal against strong SOC where the Dirac lines are made up of clean valence and conduction bands (Fig. 3c). The bandgap is 0.23 eV at X and the 0.17 eV dispersion of the DNL in 1-AL Bi is smaller than that of the 3-AL Bi. The smaller dispersion of the DNL indeed helps with standing out it, however, the states at point X do strongly interference with the DNL in 1-AL Bi, which slightly improves in 5-AL Bi but the states in Γ-Y become another issue competing with the DNL states (Fig. 3f). Although the preparation of 1-AL Bi atoms appears difficult, it is a feasible route to apply external nonsymmorphic symmetry potential field to 2D electron systems[45]. The 5-AL Bi is 19.22 meV/Bi more stable than the straightforward 3+2-AL structure (Supplementary Fig. S8), which was recently synthesized on the BP substrate [46].

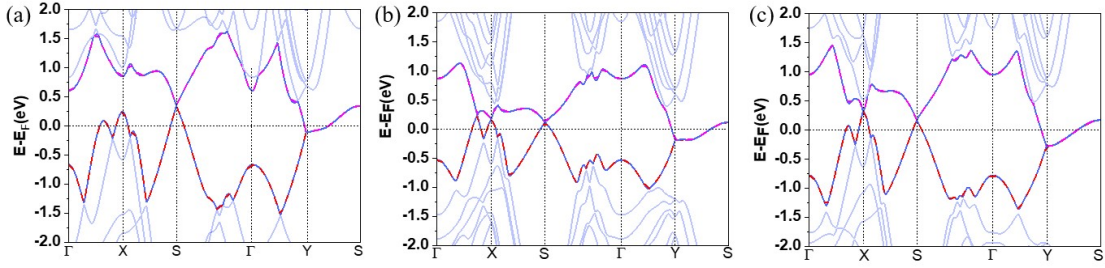

Fig. 4 Electronic band structures along high-symmetry directions of tri-layer P (a), As (b) and Sb (c) in the presence of SOC.

We last extend the 3-AL Bi to P, As and Sb 3-ALs using the same brick-wall structures. They are all DNL semimetals against strong SOC, though they have certain extra bands around the X point near the Fermi level (Figs. 4a-4c). Differently from the non-trivial one found in 3-AL Bi, the 3-AL P has no band inversion around the time-reversal invariant X point. A 3% tension strain applied along the arm-chair direction of 3-AL P lowers the VBM at the X point below the Fermi level and the DNL state still maintains (Supplementary Fig. S9). In terms of the Sb and As 3-ALs, the VB and CB are overlapped at the X point, but are usually tunable by strain. Among all these candidates showing the DNL semimetal properties against SOC, the 3-AL Bi appears the best platform for studying exotic properties purely induced by the DNL states.

In summary, we found that the nonsymmorphic space group protects the formation of 2D DNL semimetals along line ($k_x$, $\pi$, 0) whose little co-group is $\mathcal{C}_{2\nu} \times Z_2^T$. A novel allotrope, i.e. the brick phase, of group VA elemental (Bi, Sb, As and P) layers imposes such symmetry and was theoretically predicted to offer 2D DNL semimetals whose degeneracy is robust against SOC. Intriguingly, the DNL states are across the Fermi level while other bands are away from it; this results in a ``neat'' DNL state around the Fermi level. Strong signatures of such ``neat'' DNL state were recently observed in synthesized 3- and 5-AL Bi on a BP substrate by our coworkers using scanning tunneling microscopy[46]. Given this experimental observation, we feel more confident with our theoretical predictions and expect more fascinating physics and properties to be discovered in the heterojunctions and/or alloys of the brick phase layers. The IT operation is not a symmetry element of any moments in the BZ of ferromagnetic materials, but is indeed a symmetry for anti-ferromagnetic layers. It means the DNL which is robust against SOC could be also found in 2D antiferromagnetic materials, whose long-range magnetism and the consequent symmetries as well as electronic structures could be tuned by a magnetic field. This finding brings about a new perspective for the exploration of magnetic DNL semimetals. More importantly, the nonsymmorphic symmetry protection of DNL semimetals can be generalized to other systems, such as phonon, photonic or magnon bands. It is general for all 2D materials that could boost the artificially design and data mining of 2D DNL materials, as well as serve as a playground for subsequent exploring of their unique physical properties using e.g. device measurements.

**Methods.**

**DFT calculations.** DFT calculations were performed using the generalized gradient approximation for the exchange-correlation potential, a plane-wave basis, and the projector augmented wave method set as implemented in the Vienna *ab-initio* simulation package (VASP)[47-49]. The energy cutoff for plane wave was set to 650 eV, the *k*-points sampling of the first Brillouin zone is 14×14×1, generated automatically by Monkhorst-Pack method[50]. The vacuum layers of all supercells are larger than 15 Å. The structures were fully relaxed until the residual force per atom was less than 0.001 eV/Å. In structural relaxation and electronic property calculations, DFT-D3 correction method is considered with the Perdew-Burke-Ernzerhof (PBE) exchange functional (PBE-D3)[51,52]. Quantum ESPRESSO[53,54] was used in phonon dispersion spectra calculations with optB86b functional for the exchange potential. The phonon dispersion was obtained by Fourier interpolation of the dynamical matrices calculated using an 16×16×1 k-mesh and a 4×4×1 qmesh with a plane-wave energy cutoff of 65 Ry. Edge states are calculated using Wannier90[55] and WannierTools[56].

**Symmetry analysis without spin-orbit coupling.** We temporarily ignore the spin degrees of freedom. Denoting the combination symmetry of $\{C_{2x}|\ (0,\ 1/2,\ 0)\}$ and $\tilde{T}$ as $\{C_{2x}\tilde{T}|\ (0,\ 1/2,\ 0)\}$, it can be shown that the square of $\{C_{2x}\tilde{T}|\ (0,\ 1/2,\ 0)\}$ is equal to -1 for all the *k* points along the Y-S axis,

$$(x,\ y,\ z) \xrightarrow{\{C_{2x}\tilde{T}|(0,1/2,0)\}} (-x,\ y+1/2,\ z) \xrightarrow{\{C_{2x}\tilde{T}|(0,1/2,0)\}} (x,\ y+1,\ z).$$

Namely, $[\{C_{2x}\tilde{T}|\ (0,\ 1/2,\ 0)\}]^2$ is equivalent to a translation of one lattice site along *y*-direction $[\{C_{2x}\tilde{T}|\ (0,\ 1/2,\ 0)\}]^2 = (0,1,0)$.

Noticing that $C_{2x}\tilde{T}$ is anti-unitary, we note the representation of $\{C_{2x}\tilde{T}|\ (0,\ 1/2,\ 0)\}$ as $D_o(C_{2x}\tilde{T})K$, then the *k* points along the line $(k_x,\pi,0)$, we have

$$[D_o(C_{2x}\tilde{T})K]^2 = e^{-ik\cdot(0,1,0)} = -1.$$

From the Kramers theorem, we know that the dimension of the matrix $M(C_{2x}\tilde{T})$ is an even number, which means that the degeneracy along the Y-S axis is at least 2. The mechanism is similar to the Kramers degeneracy protected by time reversal symmetry satisfying $\tilde{T}^2 = -1$. The difference is that the here the degeneracy purely comes from orbital degrees of freedom instead of spin.

The product of $\{C_{2x}| (0, 1/2, 0)\}$ and $M_z$ yields $\{M_y| (0, 1/2, 0)\}$. It can be shown that $\{C_{2x}| (0, 1/2, 0)\}$ and $\{M_y| (0, 1/2, 0)\}$ are commuting,

$$(x, y, z) \xrightarrow{\{C_{2x}|(0,1/2,0)\}} (x, -y+1/2, -z) \xrightarrow{\{M_y|(0,1/2,0)\}} (x, y, -z),$$
$$(x, y, z) \xrightarrow{\{M_y|(0,1/2,0)\}} (x, -y+1/2, z) \xrightarrow{\{C_{2x}|(0,1/2,0)\}} (x, y, -z).$$

If we denote the representation matrices of $\{C_{2x}| (0, 1/2, 0)\}$ and $\{M_y| (0, 1/2, 0)\}$ as $D_o(C_{2x})$ and $D_o(M_y)$ respectively, then

$$D_o(C_{2x}) D_o(M_y) = D_o(M_y) D_o(C_{2x}).$$

This indicates that the $\{C_{2x}| (0, 1/2, 0)\}$ and $\{M_y| (0, 1/2, 0)\}$ symmetry elements have no contribution to the orbital degeneracy.

Now we consider the spin degrees of freedom, which are always degenerate if SOC is ignored. So the total degeneracy involving the spin and orbital is 4. In other words, without inclusion of SOC, the combination symmetry $\{C_{2x}\tilde{T}| (0, 1/2, 0)\}$ alone can protect the 4-fold degenerate nodal line along Y-S axis in BZ.

**Symmetry analysis with spin-orbit coupling.** If we consider spin-orbit coupling, then the spin-rotation is locked with the corresponding lattice rotation. When consider the factor systems of the projective representation, we need to consider the contributions from both the orbital and the spin. Here we consider the following three quantum numbers: $\tilde{T}^2$, $[\{C_{2x}\tilde{T}| (0, 1/2, 0)\}]^2$, and the commuting/anti-commuting relation between $\{C_{2x}| (0, 1/2, 0)\}$ and $\{M_y| (0, 1/2, 0)\}$.

1), $(\tilde{T})_o^2 = E$ in the orbital sector, and $(\tilde{T})_s^2 = -1$ in the spin-1/2 sector. The combination gives

$$\tilde{T}^2 = -1 \tag{A1}$$

for all the *k* points in the BZ. This guarantees a 2-fold Kramers degeneracy.

2), as shown in the case without SOC (subsection II), the relation $[\{C_{2x}\tilde{T}| (0, 1/2, 0)\}]_o^2 = -1$ holds for the *k* points along the Y-S axis in the orbital sector. While in the spin-1/2 sector, the double valued Rep $D_s(\{C_{2x}\tilde{T}| (0, 1/2, 0)\}) = i\sigma_z K$ yields $[\{C_{2x}\tilde{T}| (0, 1/2, 0)\}]_s^2 = 1$. Considering both the orbital and spin degrees of freedom, we still have

$$[\{C_{2x}\tilde{T}| (0, 1/2, 0)\}]^2 = -1, \tag{A2}$$

which also gives rise to a 2-fold degeneracy.

The question is if the two invariants in 1) and 2) are enough to protect the 4-fold degeneracy. This is equivalent to ask if the 4-dimensional Rep is reducible or not. If one can combine the spin and orbital bases to form a 2-dimensional irreducible Rep, then the 4-fold degeneracy is not stable against SOC. Actually, such a 2-dimensional Rep indeed exist:

$$M(\tilde{T})K = i\sigma_y K,$$
$$M(\{C_{2x}| (0, 1/2, 0)\}) = I,$$

here I is the identity matrix. From the Rep matrices it is easy to verify that $\tilde{T}^2 = [\{C_{2x}\tilde{T}| (0, 1/2, 0)\}]^2 = -1$. Therefore, We need more symmetry to guarantee the 4-fold degeneracy if SOC is present.

3) as shown in the case without SOC (subsection II), for the *k* points along the Y-S axis in the orbital sector, $\{C_{2x}| (0, 1/2, 0)\}$ commutes with $\{M_y| (0, 1/2, 0)\}$, namely $[\{C_{2x}| (0, 1/2, 0)\}]_o [\{M_y| (0, 1/2, 0)\}]_o = [\{M_y| (0, 1/2, 0)\}]_o [\{C_{2x}| (0, 1/2, 0)\}]_o$. While in the spin-1/2 sector, the double valued Rep $D_s(\{C_{2x}|(0, 1/2, 0)\}) = i\sigma_x$, $D_s(\{M_y|(0, 1/2, 0)\}) = i\sigma_y$ yields $[\{C_{2x}| (0, 1/2, 0)\}]_s [\{M_y| (0, 1/2, 0)\}]_s = - [\{M_y| (0, 1/2, 0)\}]_s [\{C_{2x}| (0, 1/2, 0)\}]_s$. Combining the two degrees of freedom, we have

$$\{C_{2x}| (0, 1/2, 0)\} \{M_y| (0, 1/2, 0)\} = -\{M_y| (0, 1/2, 0)\} \{C_{2x}| (0, 1/2, 0)\}. \quad (A3)$$

Namely, when considering SOC, the two operations $\{C_{2x}| (0, 1/2, 0)\}$ and $\{M_y| (0, 1/2, 0)\}$ anti-commute. Thus the representation of $\{C_{2x}| (0, 1/2, 0)\}$ cannot be proportional to identity matrix any more. Thus, it is impossible to construct a 2-dimensional Rep satisfying (A1), (A2), (A3). In other words, when the symmetry elements $\tilde{T}$, $\{C_{2x}| (0, 1/2, 0)\}$ and $\{M_y| (0, 1/2, 0)\}$ are included then the 4-dimensional Rep is irreducible, which guarantees the 4-fold degeneracy for the *k* points along Y-S line.

Actually, for the projective Reps of the little co-group $C_{2v} \times Z_2^T$, there is another independent quantum number $[\{M_y\tilde{T}| (0, 1/2, 0)\}]^2$. Since it is always trivial in our case, we ignored it in above discussion.

**Proof of linear dispersion of nodal line.** Suppose that the 4-dimensional irreducible projective Rep *M(g)* at k ∈ (k$_x$, π, 0) is carried by the four fermionic bases $(\psi_k^\alpha)^\dagger|vac\rangle, \alpha = 1,2,3,4$, then for $g \in G_0^k$, we have

$$g(\psi_k^\dagger)g^{-1} = \psi_k^\dagger D(g) K_{s(g)}, g\psi_k g^{-1} = K_{s(g)} D(g)^\dagger \psi_k$$

Liner Reps of $C_{2v} \times Z_2^T$ and their bases are listed in TABLE1. The degeneracy of the energy remains to be four along the high symmetry line but reduces to two away from the high symmetry line. If the dispersion is linear, then the perturbation should contain terms linear in $\delta k_y$ and $\delta k_z$, namely

$$H_k = \psi_{k+\delta k_y}^\dagger (\delta k_y \Gamma_y) \psi_{k+\delta k_y} + \psi_{k+\delta k_z}^\dagger (\delta k_z \Gamma_z) \psi_{k+\delta k_z} + v_x \delta k_x \psi_{k+\delta k_x}^\dagger \psi_{k+\delta k_x}$$

(1)

Where $v_x$ is a function of $k_x$, and $\Gamma_y, \Gamma_z$ are 4 by 4 Hermitian matrices

$$\Gamma_y^\dagger = \Gamma_y, \Gamma_z^\dagger = \Gamma_z \qquad (2)$$

Noticing that $\delta k_y, \delta k_z$ carry $B_1, B_2$ Reps of $\mathcal{C}_{2v}$, respectively,

$$C_{2x}\delta k_y = -\delta k_y, \ C_{2x}\delta k_z = -\delta k_z, \ M_z\delta k_y = \delta k_y, \ M_z\delta k_z = -\delta k_z$$

and are invariant under $\tilde{T}$

$$\tilde{T}\delta k_y = \delta k_y, \tilde{T}\delta k_z = \delta k_z$$

Accordingly, the matrices $\Gamma_y, \Gamma_z$ should vary under $h \in \mathcal{C}_{2v}$ in the same way as $k_y, k_z$:

$$M(h)\Gamma_y M(h)^\dagger = D^{(B_1)}(h)\Gamma_y, \qquad (3)$$

$$M(h)\Gamma_z M(h)^\dagger = D^{(B_2)}(h)\Gamma_z \qquad (4)$$

such that the Hamiltonian is invariant under the symmetry group. Above equations indicate that $\Gamma_y, \Gamma_z$ are the CG coefficients that combine the bases of the direct product Rep $M(h) \otimes M^*(h)$ into the $B_1, B_2$ irreducible bases, respectively. Similarly, for the anti-unitary operator $\tilde{T}$, we have $M(\tilde{T})K\Gamma_m KM(\tilde{T})^\dagger = \Gamma_m$ for m = x, y or equivalently

$$M(\tilde{T})\Gamma_m^* M(\tilde{T})^\dagger = \Gamma_m, \quad m = x, y \qquad (5)$$

If $\Gamma_m$ satisfy relations (2)-(5), then the energy splitting will be linear in $\delta k_y$ and $\delta k_z$. Therefore, the existence of linear dispersion along m-direction is equivalent to the existence of the matrix $\Gamma_m$ satisfying relations (2)-(5).

Noticing that $M(\tilde{T})^* = M(\tilde{T}), M(\tilde{T})^\dagger M(\tilde{T}) = I$ and $[M(\tilde{T})K]^2 = M(\tilde{T})^2 = -1$, so

$$M(\tilde{T})^T = M(\tilde{T})^{-1} = -M(\tilde{T})$$

meaning that $M(\tilde{T})$ is an anti-symmetric real matrix. Furthermore, it can be checked that invariants[57] $\omega(\tilde{T}, h) = \omega(h, \tilde{T}) = 1$ for $h \in \mathcal{C}_{2v}$, namely, $M(\tilde{T})KM(h) = M(h)M(\tilde{T})K$ or equivalently

$$M(\tilde{T})M(h)^* M(\tilde{T})^{-1} = M(h) \qquad (6)$$

Owing to (6), it is convenient to introduce the new basis $\tilde{\psi}_k = M(T)\psi_k$ in which the Hamiltonian (4) takes the following form

$$H_k = \sum_{m=y,z} \tilde{\psi}_{k+\delta k_m}^\dagger (\delta k_m \tilde{\Gamma}_m) \tilde{\psi}_{k+\delta k_m} + v_x \delta k_x \tilde{\psi}_{k+\delta k_x}^\dagger [M(\tilde{T})]^T \tilde{\psi}_{k+\delta k_x}$$

Here $\tilde{\Gamma}_m = \Gamma_m M(\tilde{T})^T$. According to (3)~(4), $\tilde{\Gamma}_m$ vary in the following way under the action of symmetry operation:

$$M(h)\tilde{\Gamma}_y M(h)^T = D^{(B_1)}(h)\tilde{\Gamma}_y, \qquad (7)$$

$$M(h)\tilde{\Gamma}_z M(h)^T = D^{(B_2)}(h)\tilde{\Gamma}_z \qquad (8)$$

In other words, $\tilde{\Gamma}_y, \tilde{\Gamma}_z$ are the CG coefficients that transform the bases of the direct product Rep $M(h) \otimes M(h)$ into the $B_1, B_2$ irreducible bases, respectively. Furthermore, the Hermitian condition (2) indicates $[\tilde{\Gamma}_m M(\tilde{T})]^\dagger = M(\tilde{T})^T(\tilde{\Gamma}_m^T)^* = \tilde{\Gamma}_m M(\tilde{T})$, namely,

$$M(\tilde{T})^T \tilde{\Gamma}_m M(\tilde{T}) = -(\tilde{\Gamma}_m^T)^*$$

On the other hand, the $\tilde{T}$ action (5) yields

$$M(\tilde{T})^T \tilde{\Gamma}_m M(\tilde{T}) = \tilde{\Gamma}_m^*$$

Comparing above two equations we have

$$\tilde{\Gamma}_m^T = -\tilde{\Gamma}_m \qquad (9)$$

Therefore, we have shown that the equations (2) ~ (5) are equivalent to (7) ~(9). The existence of nonzero matrices $\tilde{\Gamma}_y, \tilde{\Gamma}_z$ satisfying (7)~(9) can be checked by computing the number of times that the irreducible Reps $B_1, B_2$ appear in the reduced Rep of $M(h) \otimes M(h)$, namely,

$$a_k^{(B_1)} = \frac{1}{|\mathcal{C}_{2v}|} \sum_{h \in \mathcal{C}_{2v}} \chi^{[M \otimes M]}(h) \chi^{(B_1)}(h)^* = 3$$

$$a_k^{(B_2)} = \frac{1}{|\mathcal{C}_{2v}|} \sum_{h \in \mathcal{C}_{2v}} \chi^{[M \otimes M]}(h) \chi^{(B_2)}(h)^* = 1$$

Where $\chi^{[M \otimes M]}(h) = \text{Tr}[P_{as} M(h) \otimes M(h)]$ is the character of $h$ in the anti-symmetric direct product of Rep $[M(h) \otimes M(h)]$, $P_{as}$ is the projection operator projecting onto the anti-symmetric subspace, and $\chi^{(B_1)}(h)$ [or $\chi^{(B_2)}(h)$] is the character of $h$ in the Rep $B_1$ (or $B_2$). Since $a_k^{(B_1)}$ and $a_k^{(B_2)}$ are nonzero, the corresponding CG coefficients form the Hermitian matrix $\tilde{\Gamma}_y, \tilde{\Gamma}_z$, respectively.

In 3-dimensions, generally $\mathcal{C}_{nv} \times Z_2^T$ (with n=2,3,4,6) can protected the Dirac nodal lines. However, if we are restricted to 2-dimensional systems, since the axis of the $C_n$ lies in the plane, only the case with n=2 is the allowed little co-group symmetry. If the hopping along z-direction is very weak, the system can be considered as quasi 2-dimensional. In this case the dispersion along $k_z$-direction is nearly flat but still linear.

| $\mathscr{C}_{2v} \times Z_2^T$ | E | $C_{2x}$ | $M_z$ | $M_y$ | $\tilde{T}$ | bases |
|---|---|---|---|---|---|---|
| $A_1$ | 1 | 1 | 1 | 1 | 1 | $k_x$ |
| $B_1$ | 1 | -1 | 1 | -1 | 1 | $k_y, \Gamma_y$ |
| $A_2$ | 1 | 1 | -1 | -1 | 1 | |
| $B_2$ | 1 | -1 | -1 | 1 | 1 | $k_z, \Gamma_z$ |

TABLE I. Linear Reps of $\mathscr{C}_{2v} \times Z_2^T$ and their bases.

**Data availability:** All data needed to evaluate the conclusions in the paper are present in the paper and/or the Supplementary Information. The data that support the findings of this study are available from the corresponding authors upon request.

**Acknowledgement**

We gratefully acknowledge financial support from the Ministry of Science and Technology (MOST) of China (Grants No. 2018YFE0202700, No.2016YFA0300504, No. 2017YFA0302903, and No. 2019YFA0308603), the National Natural Science Foundation of China (Grants No. 11622437, No. 61674171, No.61761166009, No. 11574392, No. 11974421, No. 11974422, No. 11774422, and No. 11774424), the Strategic Priority Research Program of Chinese Academy of Sciences (Grant No. XDB30000000), the Fundamental Research Funds for the Central Universities, China, and the Research Funds of Renmin University of China [Grants No. 16XNLQ01, No. 19XNLG11, No. 19XNQ025(W.J.) and No. 20XNLG51 (D.P.G.), and No. 19XNLG13 (K.L.)]. P.-J. G. was supported by China Postdoctoral Science Foundation funded project (GrantNo.2020TQ0347). Calculations were performed at the Physics Lab of High-Performance Computing of Renmin University of China and Shanghai Supercomputer Center.

**Author Contributions**

W.J. conceived this project. P.G., Z.L., K.L. and Z.-Y.L. performed the symmetry analysis; D.G. and W.J performed the first principle calculations; M.F. and L.C. inspired this work with their preliminary experimental results; they, together with S.T., participated in data analysis and discussion; D.G., P.G., Z.L and W.J. wrote the manuscript with inputs from the all authors.

**Additional information**

Additional data related to this paper are available from the corresponding authors upon request.

**Competing interests:** The authors declare that they have no competing interests.

**Support Information for**
**Symmetry guaranteed Dirac-line semimetals in two-dimensions against strong spin-orbit coupling**

Deping Guo[1†], Pengjie Guo[1,2,6†], Shiliang Tan[3], Min Feng[4,5], Limin Cao[4], Zhengxin Liu[1,*], Kai Liu[1,*], Zhong-Yi Lu[1] and Wei Ji[1,*]

[1]*Beijing Key Laboratory of Optoelectronic Functional Materials & Micro-Nano Devices, Department of Physics, Renmin University of China, Beijing 100872, China*
[2] *Songshan Lake Materials Laboratory, Dongguan, Guangdong 523808, China*
[3]*Hefei National Laboratory for Physical Sciences at the Microscale, University of Science and Technology of China, Hefei, Anhui 230026, China*
[4]*School of Physics and Technology and Key Laboratory of Artificial Micro- and Nano-Structures of Ministry of Education, Wuhan University, Wuhan 430072, China*
[5]*Institute for Advanced Studies, Wuhan University, Wuhan 430072, China*
[6]*Beijing National Laboratory for Condensed Matter Physics, Institute of Physics, Chinese Academy of Sciences, Beijing 100190, China*
[†]*These authors contributed equally to this work.*

*Corresponding authors. Email: liuzxphys@ruc.edu.cn (Z.L.), kliu@ruc.edu.cn (K.L.) and wji@ruc.edu.cn (W.J.);

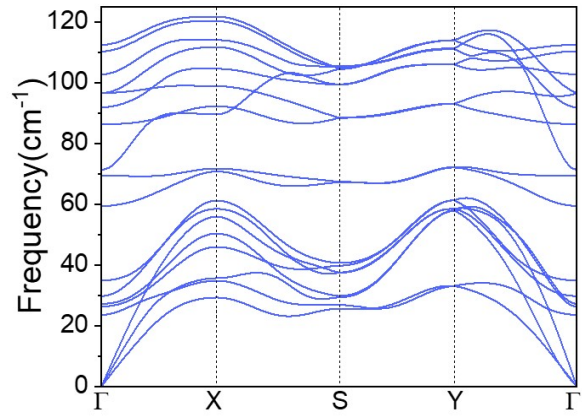

Supplementary Figure1. Phonon dispersion spectra of the Bi brick phase.

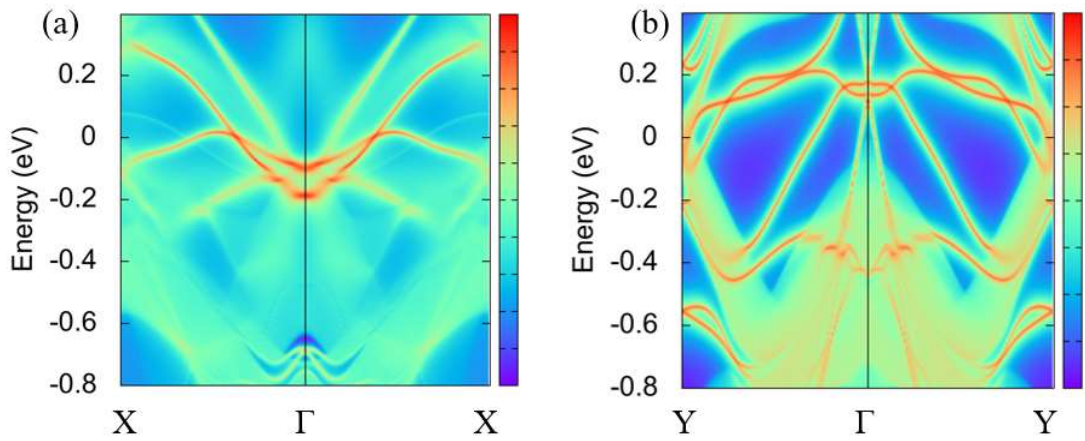

Supplementary Figure 2. Edge states along the (010) (a) and (100) (b) directions of the Bi brick phase.

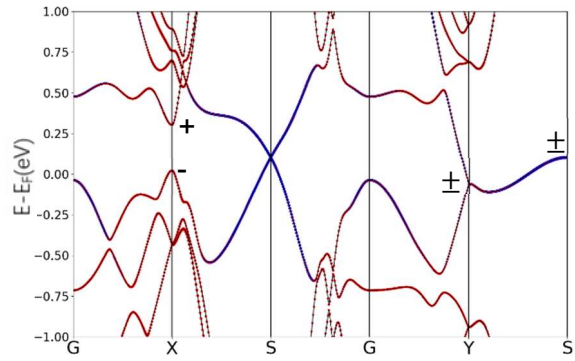

Supplementary Figure 3. Orbital-resolved band structures of the Bi brick phase. The blue lines indicate the strength of the $p_z$ components and the red lines represent the $p_x$ and $p_y$ components. Symbols ``+'' and ``−'' represent the parities of wavefunctions

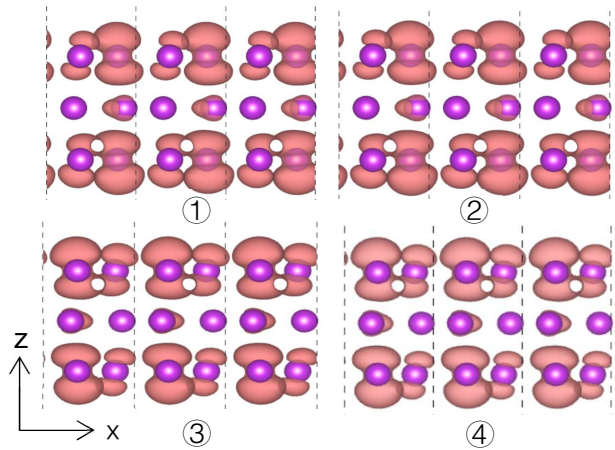

Supplementary Figure 4. Plots of partial charge density (PCD) of the quadruple-degenerated states at the S point. The isosurface value is 0.001 $e$/Bohr$^3$.

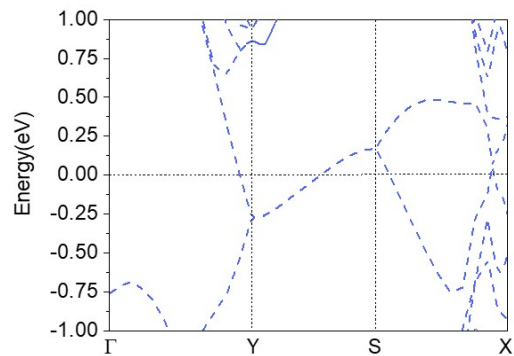

Supplementary Figure 5. Electronic band structures along high-symmetry directions of the Brillion Zone of the Bi brick phase predicted using the HSE06 functional without SOC.

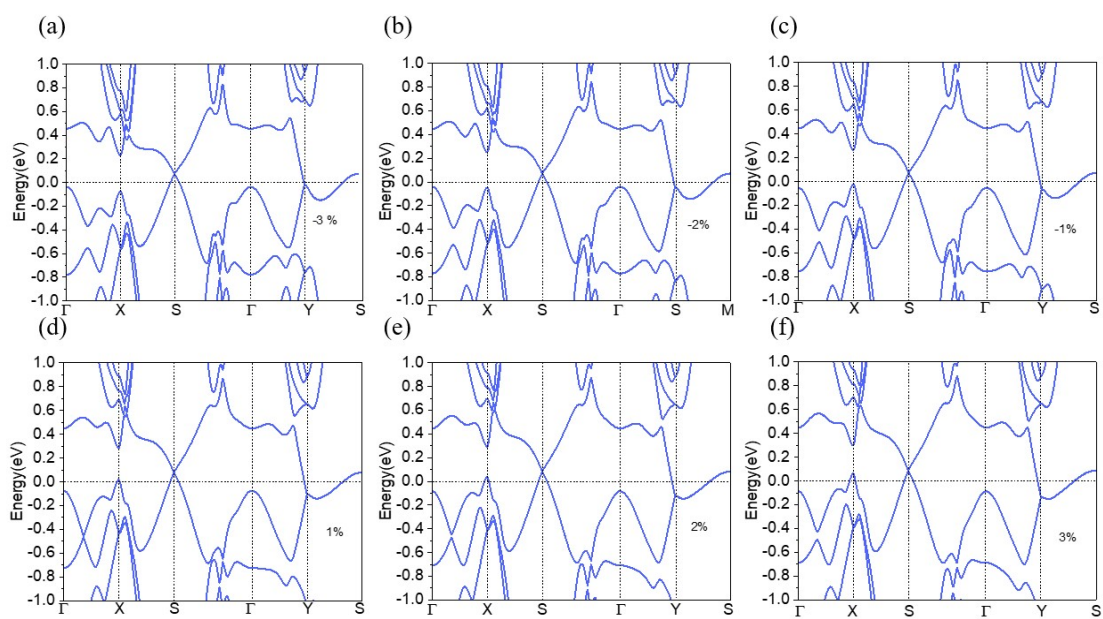

Supplementary Figure 6. Electronic band structures along high-symmetry directions of the Brillion Zone of the Bi brick phase, under uniaxial strain applied along the x direction from -3% (a) to 3% (f) with a step of 1%.

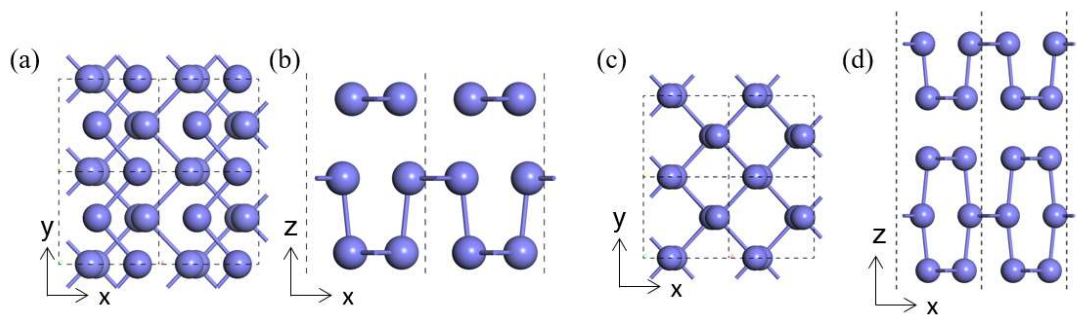

Supplementary Figure 7. Top (a, c) and side (b, d) views of the geometric structures of the 2+1 AL (a, b) and 3+2 AL (c, d) Bi layers in comparison with the 3- and 5-AL Bi layers.

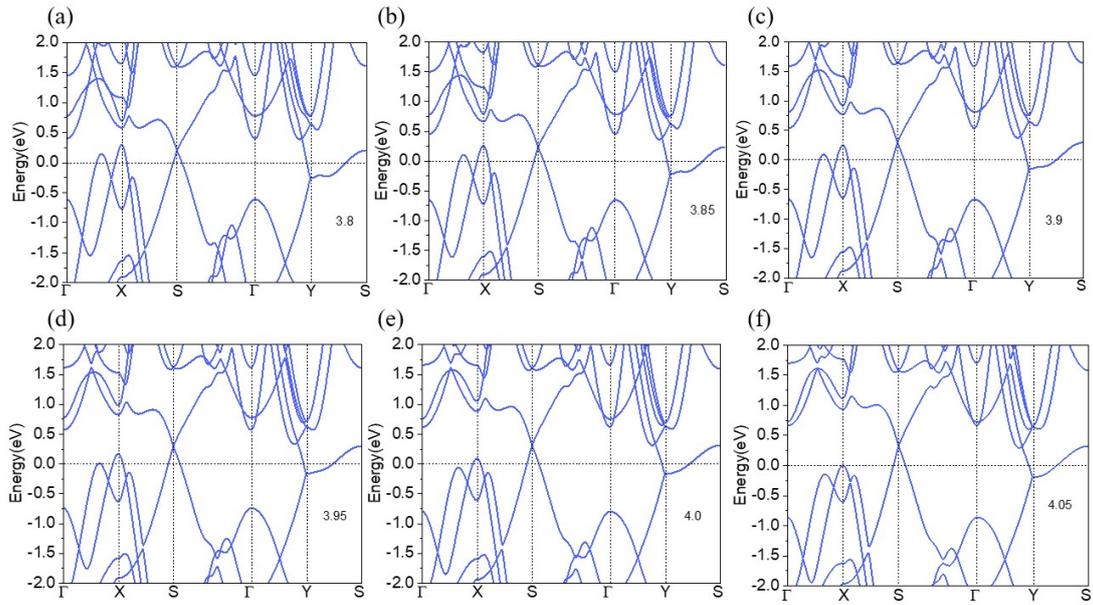

Supplementary Figure 8. Band structures of applying different strain to the tri-layer P along the arm-chair direction by tuning crystal lattice parameters from 3.80 (a) to 4.05 (f) Å with a step of 0.05 Å. While the equilibrium lattice constant is 3.93 Å, the 4.05 Å lattice, effectively with a ~3% tensile strain applied, moves the highest VB state at the X point below the Fermi level.